\documentclass[a4paper,11pt]{article}
\pdfoutput=1 

\usepackage{jinstpub} 

\usepackage{lineno}
\RequirePackage{comment}
\RequirePackage{physics}
\title{\boldmath Customized calibration sources in the JUNO experiment}

\author[a,1]{A. Takenaka,\note{Corresponding author.}}
\author[a]{J. Hui,}
\author[b]{R. Li,}
\author[b]{S. Hao,}
\author[b]{J. Huang,}
\author[b]{H. Lai,}
\author[b]{Y. Li,}
\author[a, b]{J. Liu,}
\author[b]{Y. Meng,}
\author[b]{Z. Qian,}
\author[b]{H. Wang,}
\author[b]{Z. Xiang,}
\author[c]{Z. Yuan,}
\author[b]{Y. Yun,}
\author[b]{F. Zhang,}
\author[a]{T. Zhang,}
\author[a]{and Y. Zhang}

\affiliation[a]{Tsung-Dao Lee Institute, Shanghai Jiao Tong University, Shanghai, China}
\affiliation[b]{School of Physics and Astronomy, Shanghai Jiao Tong University, Shanghai, China}
\affiliation[c]{Institute of Modern Physics, Fudan University, Shanghai, China}

\emailAdd{akira.takenaka@sjtu.edu.cn}

\abstract{
We customized a laser calibration system and four radioactive $\gamma$-ray calibration sources for the Jiangmen Underground Neutrino Observatory (JUNO), a 20-kton liquid scintillator-based neutrino detector.
The laser source system was updated to realize the isotropic light emission timing within $\pm0.25$~nsec level and to allow the tuning of the laser intensity covering more than four orders of magnitude. 
In addition, methods to prepare four different radioactive sources ($^{18}{\rm F}$, $^{40}{\rm K}$, $^{226}{\rm Ra}$, and $^{241}{\rm Am}$), covering energies from $\order{10}$~keV to $\order{1}$~MeV, for the JUNO detector were established in this study.
The radioactivity of each source and the risk of radioactive substance leaking into the detector from the source were confirmed to meet the experimental requirements.
}


\arxivnumber{2410.01571} 

\begin{document}
\maketitle
\flushbottom

\section{Introduction}
\label{sec:intro}
The Jiangmen Underground Neutrino Observatory (JUNO) experiment~\cite{JUNO:2021vlw} is currently under construction in Jiangmen, China, and will be the world's largest liquid scintillator detector. 
The primary objective of the experiment is to conduct a comprehensive measurement of neutrino oscillation parameters, including the neutrino mass ordering, by precisely measuring the energy spectrum of reactor neutrinos. 
In addition, the experiment will observe atmospheric neutrinos, solar neutrinos, geoneutrinos, and neutrinos from supernovae, and will explore new physics through other exotic searches. 
The detector consists of a spherical acrylic vessel containing a 20-kiloton liquid scintillator region, supported by a stainless steel structure. 
Two types of photomultiplier tubes (PMTs) will be installed on the stainless steel structure; 17,612 20-inch PMTs and 25,600 3-inch PMTs.
The stainless steel structure, along with the outer side of the acrylic vessel and PMTs, will be submerged in pure water.
Scintillation and Cherenkov light will be emitted from charged particles generated by neutrino interactions in the liquid scintillator and detected by the PMTs. 
These signals will be processed by a dedicated electronics system. 
The responses of the liquid scintillator, PMTs, and electronics will be calibrated using laser light sources and various $\gamma$-ray sources. \par
To achieve the primary objective of determining the neutrino mass ordering, it is essential to keep the uncertainty in the energy scale below 1\%~\cite{Zhan:2009rs, Li:2013zyd}, making the calibration system of the experiment critical. 
The JUNO detector will be equipped with multiple calibration source deployment devices, each covering different regions of the detector according to their mechanical characteristics~\cite{JUNO:2020xtj, Hui:2021dnh, Guo:2019fkf, Zhang:2020grf, Feng:2018xad}. 
The calibration sources introduced in this paper will be deployed within the detector using these devices. 
The laser system will be used to calibrate the timing response of the PMTs and the non-linearity of the PMT and electronics responses.
Based on the basic laser system design presented in the previous paper~\cite{Zhang:2018yso}, we updated the shape of the diffuser ball, which is attached to the tip of the laser light source, and the light intensity tuner.
This is discussed in detail in Section~\ref{sec:laser}.
The non-linearity of the light output of the liquid scintillator as a function of energy in the reactor neutrino energy range will be calibrated using various radioactive sources~\cite{JUNO:2020xtj}.
In this study, we developed two new radioactive sources: fluorine-18 ($^{18}{\rm F}$, $\beta^{+}$-decay) and potassium-40 ($^{40}{\rm K}$, 1.46~MeV $\gamma$-ray). 
Additionally, following the recent development of a new trigger system that can lower the detector energy threshold to approximately 20~keV to enhance sensitivity for neutrino observations from astrophysical sources~\cite{Ye:2021zso, Morton-Blake:2023ria}, radium-226 ($^{226}{\rm Ra}$, 186~keV $\gamma$-rays) and americium-241 ($^{241}{\rm Am}$, 59.5~keV $\gamma$-rays) sources were developed to calibrate the energy range around and below 100~keV in the JUNO detector.
The radioactivities of these sources were estimated using germanium- and silicon-based detectors to confirm that they satisfy the experimental needs. 
Section~\ref{sec:customsource} of this paper provides a detailed description of the development of these radioactive sources, before the conclusion in Section~\ref{sec:conclusion}.



\section{Updates of laser calibration system}
\label{sec:laser}
The laser system in the JUNO experiment serves the following two primary purposes:
\begin{enumerate}
\item Calibration of response time offsets between different PMTs:
The detection time of photoelectrons is expected to vary across channels due to differences in the length of the readout cable at each PMT, the transit time of photoelectrons within the PMT bulbs, and the response time of each electronic channel. 
In the JUNO detector, the position of charged particles is expected to be reconstructed with a spatial resolution of approximately 10~cm using photon detection times from the PMTs~\cite{Huang:2022zum}, requiring that response time offsets between PMTs be calibrated to a sub-nanosecond level. 
Additionally, the laser system helps to monitor any evolution of PMT responses over time.
\item Calibration of non-linearities in the 20-inch PMT and the associated electronics responses: 
To meet the requirement of keeping the energy scale uncertainty below 1\%, it is essential to calibrate the non-linearities in the 20-inch PMT and electronics responses. 
As mentioned earlier, the JUNO experiment will place 3-inch PMTs in the space between the 20-inch PMTs. 
By illuminating both the 20-inch (charge mode) and 3-inch PMTs (mostly single photoelectron counting mode) with laser light of various intensities and comparing their responses, the non-linearity of the output from the 20-inch PMT system can be calibrated~\cite{JUNO:2020xtj}. 
In reactor neutrino events in JUNO, the number of photoelectrons detected by each 20-inch PMT can range from a single photoelectron (typical mean light luminosity $\order{0.01}$~photoelectron) to the equivalent of over 100 photoelectrons. 
The laser system must cover this range of intensity and also be capable of varying the intensity over four orders of magnitude.
\end{enumerate}
The basic configuration of the laser system remains unchanged from what was reported in the previous publication~\cite{Zhang:2018yso}. 
A laser emitting ultraviolet (UV) light with a wavelength of 266~nm (FQSS266-Q4-1k~\cite{UVLaser}) will be used to produce photons similar in wavelength range and temporal structure to the scintillation light emitted by charged particles through the absorption and emission processes described below.
The pulse width is estimated to be less than 1~nsec (full width at half maximum)~\cite{UVLaser}, which is shorter than the time resolution of the PMTs installed in the detector.
The laser light will be split into two paths: one directed toward the JUNO detector and the other toward an independent PMT system outside the detector, which will measure the laser emission time and produce a trigger signal to the JUNO data-taking system.
The laser light will be transmitted to the detector through a 50-meter optical fiber, with a diffuser ball attached to the fiber end to diffuse the light across all PMTs inside the detector. 
The diffuser ball is made of polytetrafluoroethylene (PTFE) and will be positioned at the center of the JUNO detector using the ACU system~\cite{Hui:2021dnh}, one of the calibration source deployment devices. 
In the JUNO liquid scintillator, photons with wavelengths shorter than approximately 400~nm are absorbed, and longer-wavelength photons are subsequently emitted by its fluor and wavelength shifter~\cite{JUNO:2020bcl}.
The UV photons emitted from the diffuser ball will undergo the same processes and re-emission photons will eventually be detected by the PMTs.
In this study, the shape of the diffuser ball was optimized to ensure uniform timing of ultraviolet light emission from its surface, and a filter was installed in the laser light path to tune the laser light intensity.

\subsection{Optimization of the diffuser ball shape}
\label{subsec:laser1}
\begin{figure}[htbp]
\centering
\includegraphics[width=0.5\linewidth, trim=2 2 2 2,clip]{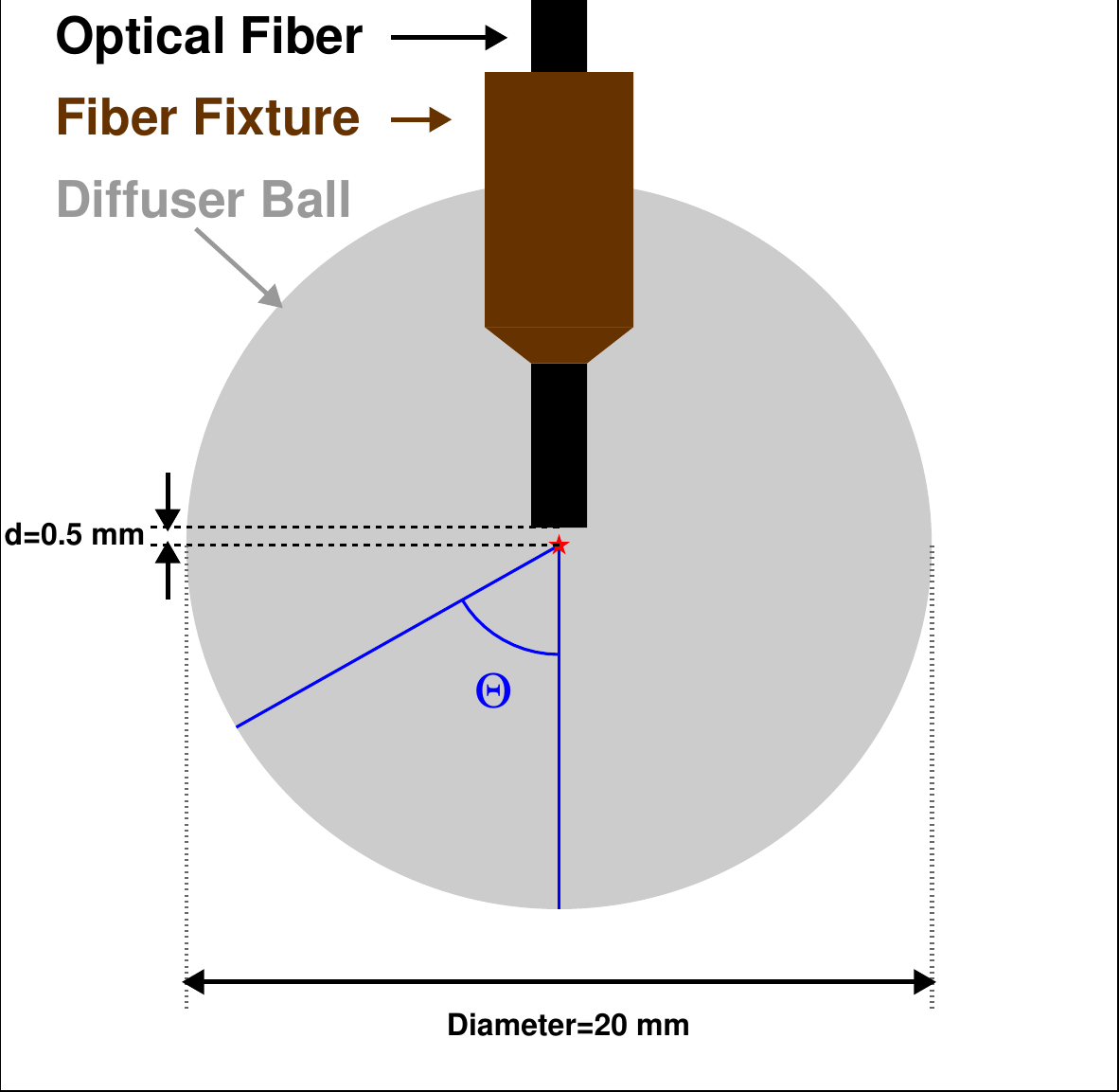}
\caption{
A cross-sectional schematic view of the diffuser ball (gray) attached to the tip of the fiber (black) and its fixture (brown).
The red star mark represents the center of the diffuser ball.
The position of fiber tip is placed offset from the center of the diffuser ball, in the direction opposite to the laser incidence, which is downward in the figure.
This offset ($d$) was optimized to be 0.5~mm as a result of the measurements presented in the text.
$\Theta$ represents the zenith angle of the diffuser ball.
}
\label{fig:laser1}
\end{figure}

The optical fiber guiding the laser light and the diffuser ball are fixed in place using a plastic (polyetheretherketone, PEEK) fixture, as shown in Figure~\ref{fig:laser1}. 
Although the laser light travels in a nearly straight line when it reaches the diffuser ball, it undergoes multiple scatterings within the ball, causing the direction of the photons to become randomized. 
As a result, the photons are eventually emitted from various angles of the ball. 
In this study, the uniformity of the light intensity and emission timing from the diffuser ball was measured using a dedicated measurement system at Shanghai Jiao Tong University.
The diffuser ball and the plastic fixture possess a symmetrical geometric structure in the azimuthal direction, and no significant non-uniformity was observed in this direction. 
Therefore, the discussion below will focus on the uniformity in the zenithal ($\Theta$ in Figure~\ref{fig:laser1}) direction of the diffuser ball. \par

\begin{figure}[htbp]
\centering
\includegraphics[width=0.5\linewidth]{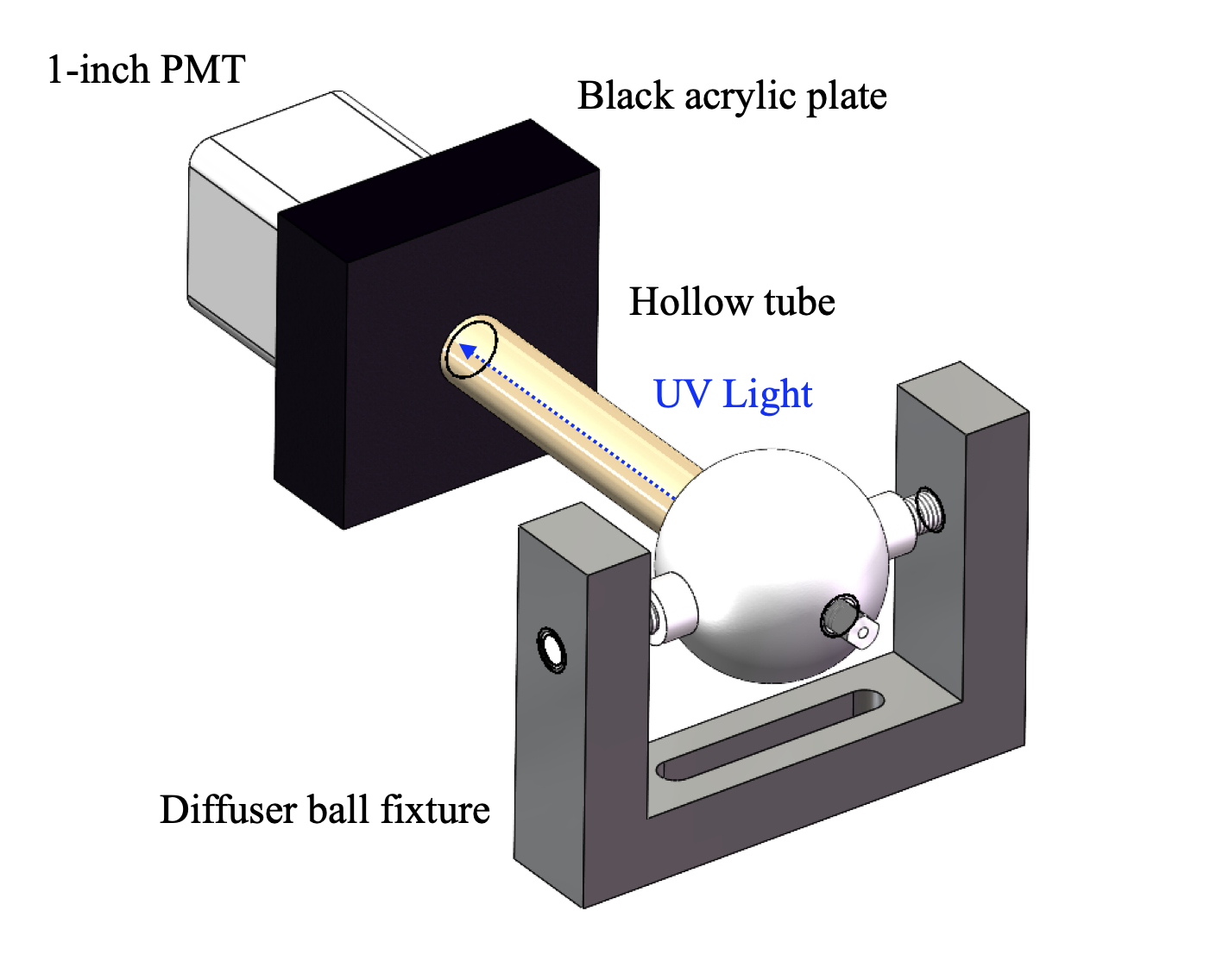}
\caption{
Schematic diagram of the setup for measuring the uniformity of light emission time and light intensity from the diffuser ball. 
The diffuser ball is held in place by a fixture, allowing only rotation.
By detecting only the light that passes through a hollow tube attached to a black acrylic plate, placed in front of the 1-inch PMT's photocathode, the relative emission time and light intensity from specific angles on the ball surface were measured.
}
\label{fig:laser4}
\end{figure}

The measurement system consists of a fast-response and UV-sensitive 1-inch PMT (R8520-406 from Hamamatsu Photonics K.K.) and an oscilloscope with waveform recording capabilities (WaveSurfer 104MXs-B from Teledyne LeCroy). 
Using the fixture and hollow tube shown in Figure~\ref{fig:laser4}, the position of the diffuser ball was fixed, and by rotating the ball relative to the PMT photocathode, the relative emission time and light intensity from various angles on the ball surface were measured. 
The signals from the PMT were sent to the oscilloscope, which was triggered by the synchronized pulse generated by the laser device.
The light intensity detected by the PMT was pre-set to a single photoelectron level. 
The light intensity from different zenithal angles of the diffuser ball was defined as the ratio of photon detections by the PMT to the number of laser emissions, while the photon emission time was defined as the average signal detection time recorded by the oscilloscope. \par

The uniformity of the light intensity and emission timing from the diffuser ball correlates with both the diameter of the ball and the position of the fiber tip within the ball. 
A larger ball diameter increases the number of photon scatterings inside the ball, enhancing uniformity; however, if the ball is too large, the impact of light absorption within the ball also increases, making it difficult to achieve the required light intensity for the experiment.
Additionally, because the laser light remains directionally biased upon reaching the ball, setting the fiber tip at the center of the ball does not completely eliminate the initial directional bias of the laser light, leaving some non-uniformity in both light intensity and photon emission timing. 
Therefore, as shown in Figure~\ref{fig:laser1}, the fiber tip is placed at a position offset from the center in the direction opposite to the direction of the incoming laser light. 
In this study, various measurements were conducted by altering the diameter of the ball and the position of the fiber tip inside it to optimize the shape of the ball. \par
\begin{figure}[htbp]
\centering 
\includegraphics[width=.75\textwidth]{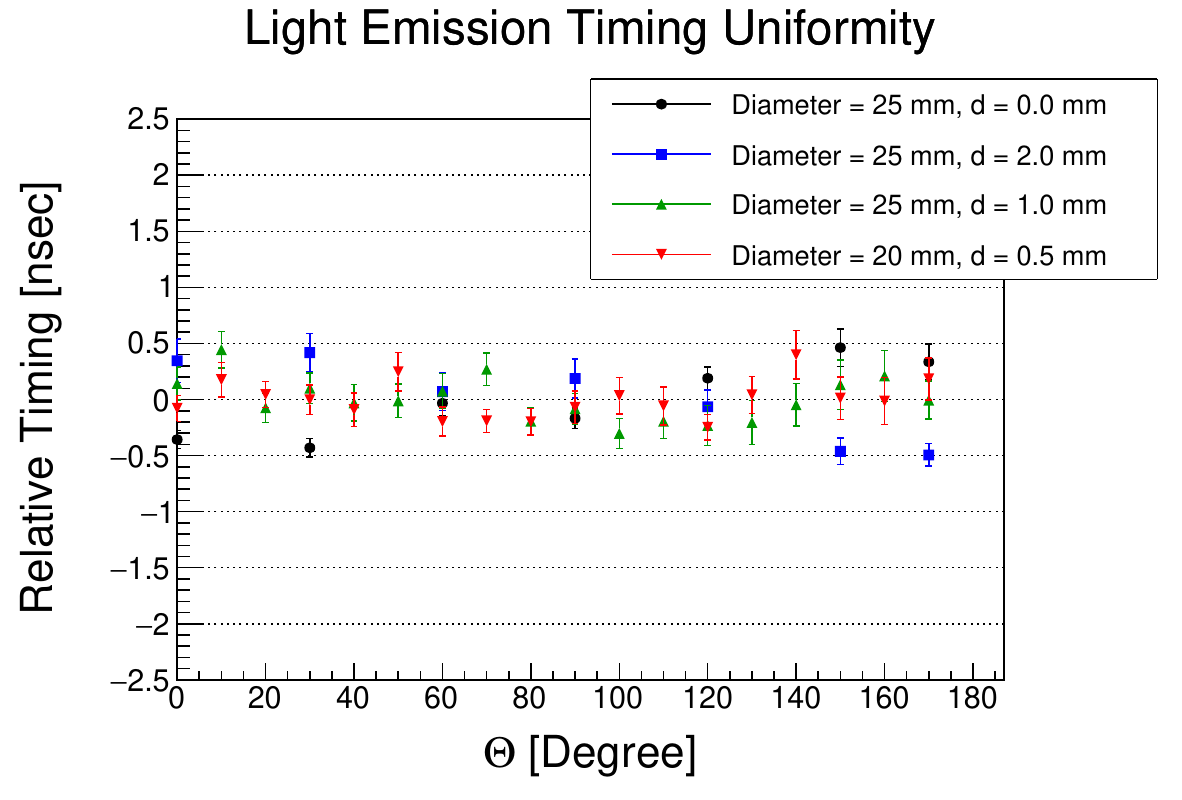}
\qquad
\includegraphics[width=.75\textwidth]{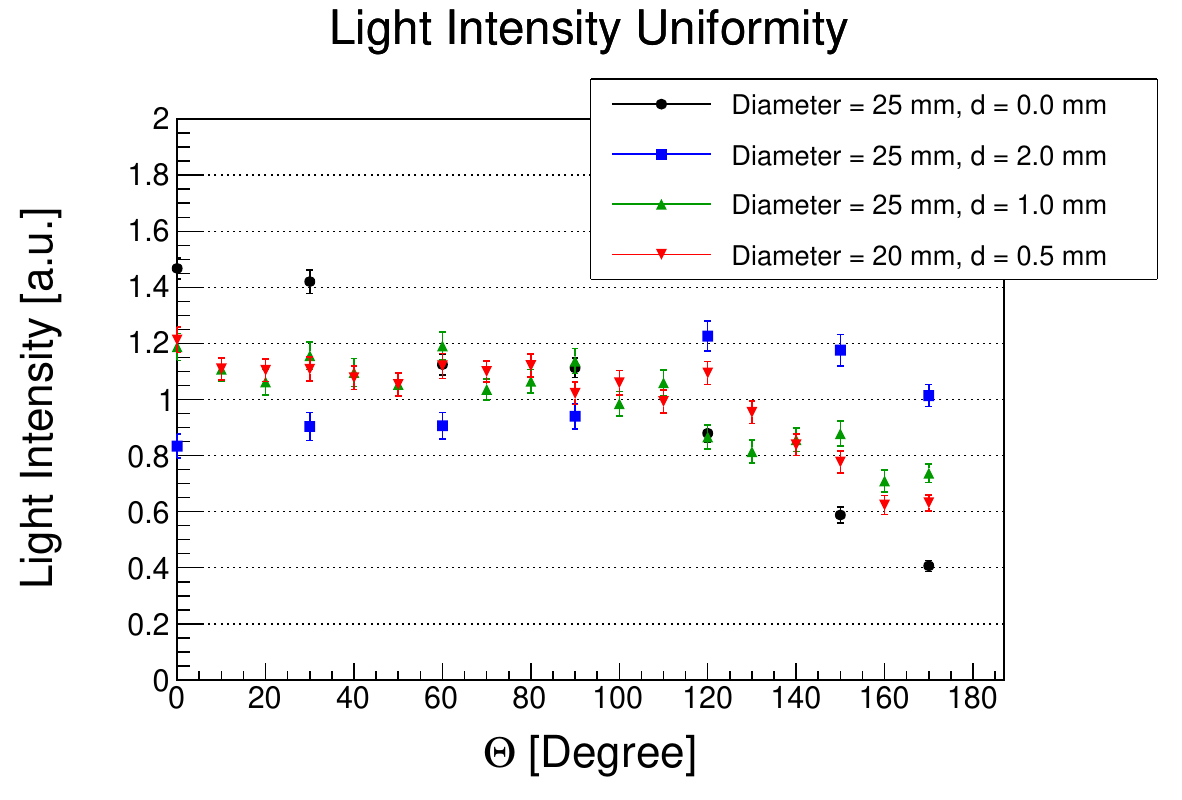}
\caption{
The top plot shows the light emission timing against the light emission direction $(\Theta)$, defined in Figure~\ref{fig:laser1}.
The Y-axis value of each data point represents the difference from the average value of all data points.
The bottom plot shows the light intensity as a function of light emission angle $(\Theta)$.
The Y-axis value of each data point represents the ratio to the average value of all data points.
For both of the plots, vertical error bars denote the statistical uncertainty.
For the balls exhibiting significant non-uniformity during the measurements taken at 30-degree intervals (plotted in black circles, blue squares), the measurement process was terminated at that stage, resulting in fewer data points compared to the optimal balls (plotted in green up-pointing triangles and red down-pointing triangles). 
The optimal balls have most data points within $\pm$0.25~nsec.
As mentioned in the text, when the fiber tip is positioned at the center of the ball ($d=0$~mm), more light is emitted more rapidly in the direction of laser incidence (small $\Theta$).
}
\label{fig:laser2}
\end{figure}

The shape of the ball was optimized based on the uniformity of photon emission timing from its surface. 
Initially, measurements were taken at seven angles ($\Theta=0$, 30, 60, 90, 120, 150, and 170~degrees). 
For balls where significant non-uniformity was detected, measurements were concluded at that stage. 
For the remaining balls, non-uniformity measurements were further conducted at 10-degree intervals.
As a result, the optimal ball shapes were determined to be a ball with a diameter of 25~mm with the fiber tip positioned 1.0~mm from the center, and a ball with a diameter of 20~mm with the fiber tip positioned 0.5~mm from the center. 
Considering the light absorption inside the ball due to increased diameter, the 20~mm diameter ball is planned for use in the JUNO experiment.
The left plot of Figure~\ref{fig:laser2} shows the dependence of the average photon emission time on the zenith angle $(\Theta)$ for various ball shapes.
For the optimal balls, the average photon emission time was uniform within $\pm0.25$~nsec across most of the angles.
The right plot of Figure~\ref{fig:laser2} shows the relationship between light intensity and zenith angle, revealing that even for the optimal ball, a non-uniformity of up to $\pm$40\% was observed.
A decrease in light intensity is generally observed on the side of the plastic fixture, which is considered to be due to the absorption of scattered light within the ball by the fixture. 
However, the non-uniformity measured in this study refers to the light intensity on the surface of the ball. 
In the JUNO detector, as previously mentioned, the UV light is absorbed by the liquid scintillator, and photons of longer wavelengths are re-emitted.
The direction of the re-emitted photons is isotropic from the point of absorption, significantly reducing the original non-uniformity on the ball surface. 
The JUNO detector simulations~\cite{Lin:2022htc} based on Geant4~\cite{GEANT4:2002zbu, Allison:2016lfl}, incorporating this non-uniformity in photon emission position, estimated that the light intensity variation observed on the PMT mounting surface, approximately 19.4 m away from the detector center (diffuser ball location), is a few percent, including the shadowing effect cast by the plastic fixture.

\subsection{Introduction of the light intensity tuner}
\label{subsec:laser2}
\begin{figure}[htbp]
\centering 
\includegraphics[width=.47\textwidth]{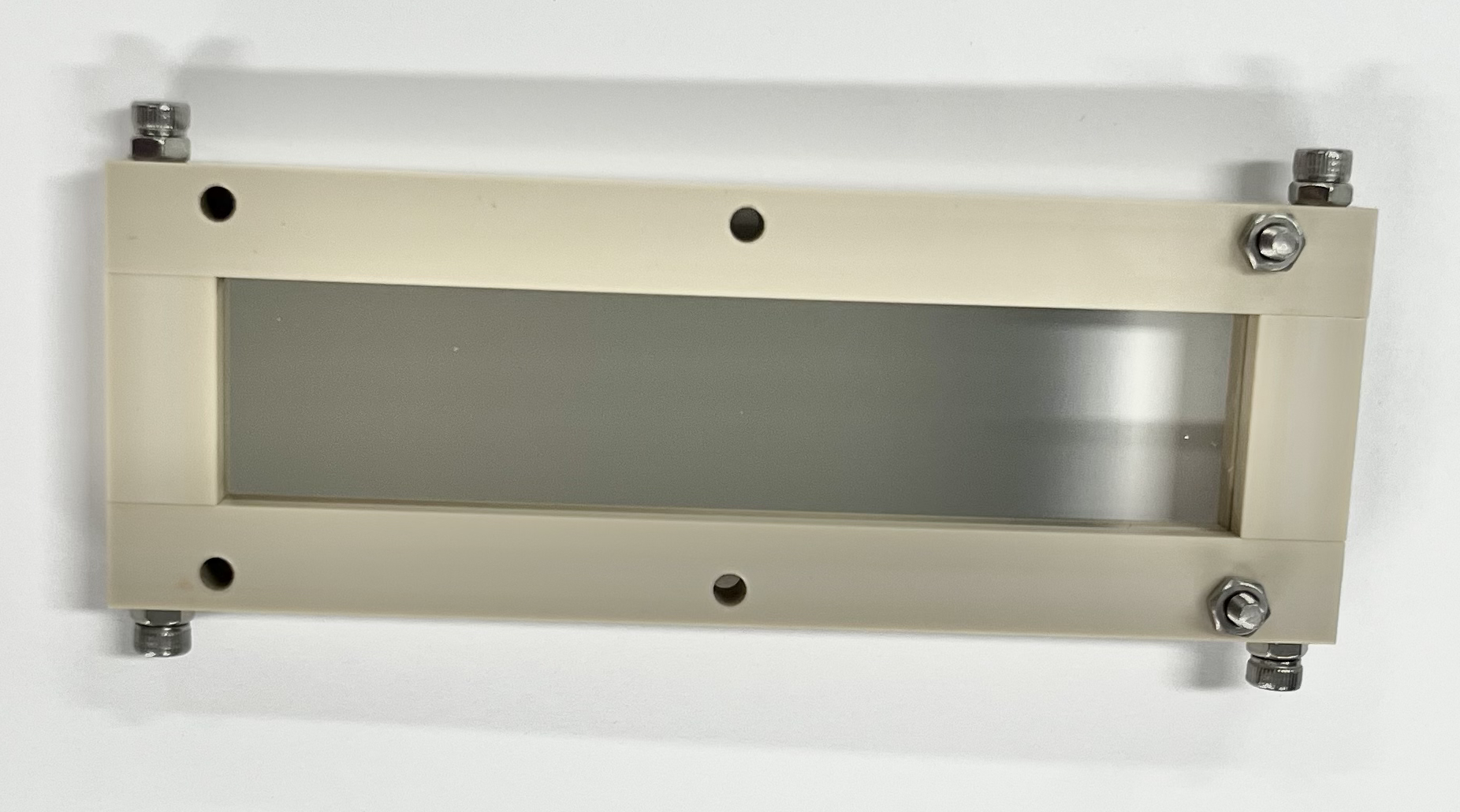}
\qquad
\includegraphics[width=.47\textwidth]{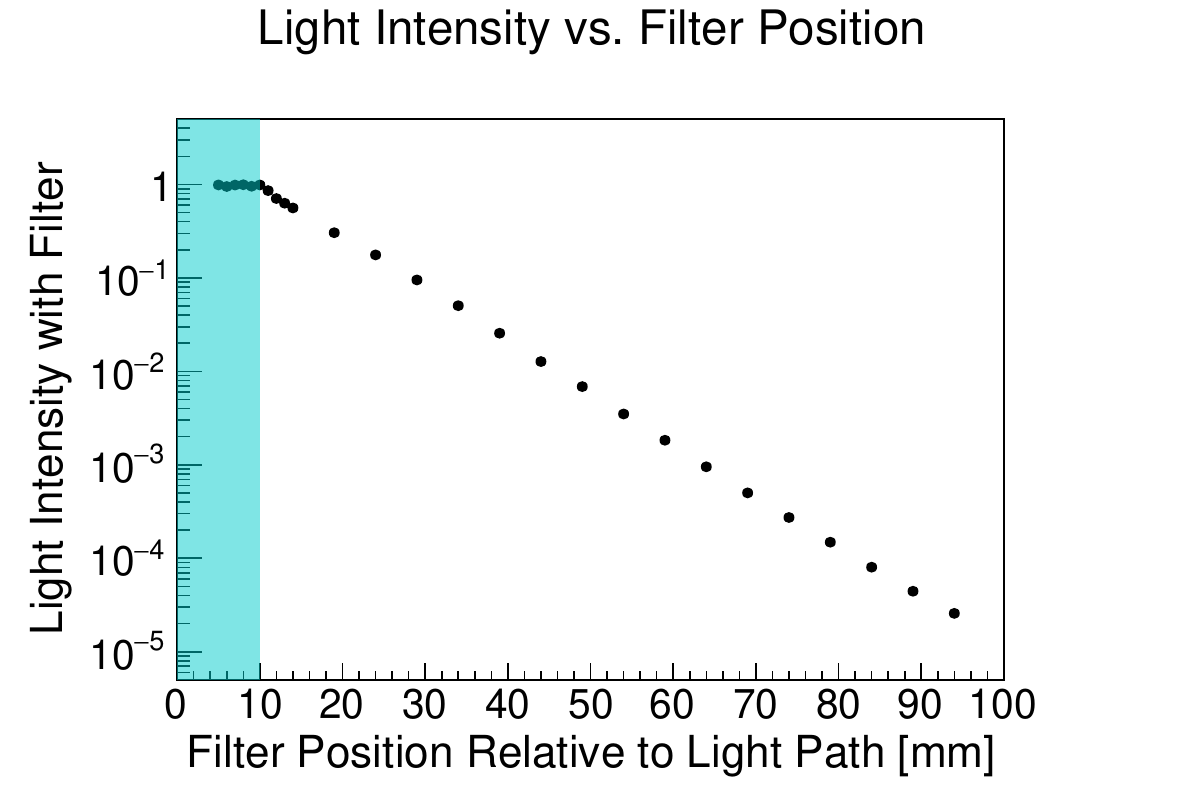}
\caption{
The left photograph shows the neutral density filter employed to tune the light intensity.
The filter is housed within plastic fixtures.
Of the 10~cm lateral length of the filter, 9~cm consists of a natural density filter.
The right plot shows the light intensity change against the change of the filter position relative to the laser path.
The Y-axis value is the ratio of the light intensity with the filter to that without.
The first 1~cm (shaded in light blue) part in the plot corresponds to the glass section where the transmitted light intensity remains unchanged.
}
\label{fig:laser3}
\end{figure}

By installing a continuously variable neutral density filter (NDL-25C-4 from Thorlabs~\cite{Filter}) in the optical path of the laser system, light intensity entering the JUNO detector is tuned over a range of more than four orders of magnitude.
As shown in the left image of Figure~\ref{fig:laser3}, this neutral density filter has a rectangular shape with a lateral length of 10~cm. 
The intensity of transmitted light varies depending on the position where the light beam passes horizontally through the filter.
The filter is mounted on an automatically controlled rail (RCP6-SA4R from IAI~\cite{Rail}), allowing its position to be adjusted relative to the fixed laser beam path, thereby enabling the control of light intensity. 
Tests of the light intensity tuning using the previously described 1-inch PMT and oscilloscope measurement system were conducted. 
The right panel of Figure~\ref{fig:laser3} illustrates the relationship between the position of the laser beam on the filter surface and the intensity of transmitted light. 
Of the total 10~cm length, 9~cm serves as the filter section where the transmitted light intensity exponentially varies, consistent with the datasheet provided by the filter manufacturer~\cite{Filter}.
This measurement also confirmed that light intensity could be tuned over a range of more than four orders of magnitude.

\section{Customized radioactive sources}
\label{sec:customsource}
As outlined in the detector calibration strategy paper~\cite{JUNO:2020xtj}, several radioactive sources were already prepared for the JUNO experiment, intended for calibrating the non-linearity and non-uniformity of the detector responses in the reactor neutrino energy range. 
These radioactive sources were acquired through commercial vendors.
In this study, new $^{18}{\rm F}$ and $^{40}{\rm K}$ sources were developed as additional radioactive sources expected to serve similar calibration purposes (see Sections~\ref{subsec:f18} and~\ref{subsec:k40}). 
Among the existing sources, the lowest energy source is the 662~keV $\gamma$-ray emission from the cesium-137 ($^{137}{\rm Cs}$) source. 
However, with the recent development of a new trigger system capable of lowering the energy threshold of the JUNO detector to approximately 20~keV, $^{226}{\rm Ra}$ and $^{241}{\rm Am}$ sources were developed to cover even lower energy ranges. 
These low-energy sources are also reported in Section~\ref{subsec:ra226am241}.
Of the newly developed sources, the $^{40}{\rm K}$, $^{226}{\rm Ra}$ and $^{241}{\rm Am}$ sources were sealed in custom-made containers.
This paper also discusses the welding techniques used during the sealing of these sources and the evaluation of the welding quality in Section~\ref{subsec:sourceweld}.

\subsection{Radioactive fluorine-18 source}
\label{subsec:f18}
Among the existing radioactive sources, the germanium-68 $(^{68}{\rm Ge})$ source undergoes electron capture, decaying into the daughter nuclide gallium-68 $(^{68}{\rm Ga})$, which then decays via $\beta^{+}$-decay into stable zinc-68 $(^{68}{\rm Zn})$ with a probability of about 89\%~\cite{McCutchan:2012xmj}. 
The positrons emitted during this $\beta^{+}$-decay annihilate within a dedicatedly designed source container made of stainless steel and PTFE~\cite{Zhang:2021tob}, resulting in the emission of two 511~keV $\gamma$-rays. 
However, the half-life of the $^{68}{\rm Ge}$ source is approximately 270~days, which is relatively short compared to the anticipated 20-year operational period of the JUNO experiment.
Since it is costly to procure the $^{68}{\rm Ge}$ source regularly, we sought a cost-effective way to repeatedly regenerate a $\beta^{+}$-decay source before each deployment at the JUNO site.
$^{18}{\rm F}$ can be generated by irradiating stable $^{19}{\rm F}$ with fast neutrons, leading to an inelastic scattering reaction, ${\rm ^{19}F(n, 2n)^{18}F}$~\cite{F18prod}. 
Subsequently, $^{18}{\rm F}$ undergoes $\beta^{+}$-decay, emitting positrons as it decays into oxygen-18 $(^{18}{\rm O})$~\cite{Tilley:1995zz}. 
These positrons then annihilate within the fluorine compound, producing two 511~keV $\gamma$-rays, similar to the $^{68}{\rm Ge}$ source. 

\subsubsection{Production test of fluorine-18}
\label{subsubsec:f181}
This study utilized a deuterium-tritium neutron generator manufactured by Northeast Normal University~\cite{Li2020, LI2020111385}. 
This neutron generator is a portable, cylindrical device with a diameter of 8.6~cm and a length of 89~cm. 
Inside the neutron generator, deuterium and tritium ions are produced and accelerated, leading to their collision with a titanium target doped with deuterium and tritium.
This triggers the deuterium-tritium reaction, producing neutrons with an energy of 14~MeV. 
Since the ${\rm ^{19}F(n, 2n)^{18}F}$ reaction requires neutrons with an energy of at least 10.4~MeV~\cite{F18prod}, neutrons from the deuterium-tritium fusion are suitable for this purpose.
The neutron yield depends on the ion source current and ion acceleration voltage, which are controllable by the user, and is estimated to be on the order of $10^{7}$ to $10^{8}$ neutrons per second. \par
\begin{figure}[htbp]
\centering 
\includegraphics[width=.6\textwidth, trim=2 2 2 2,clip]{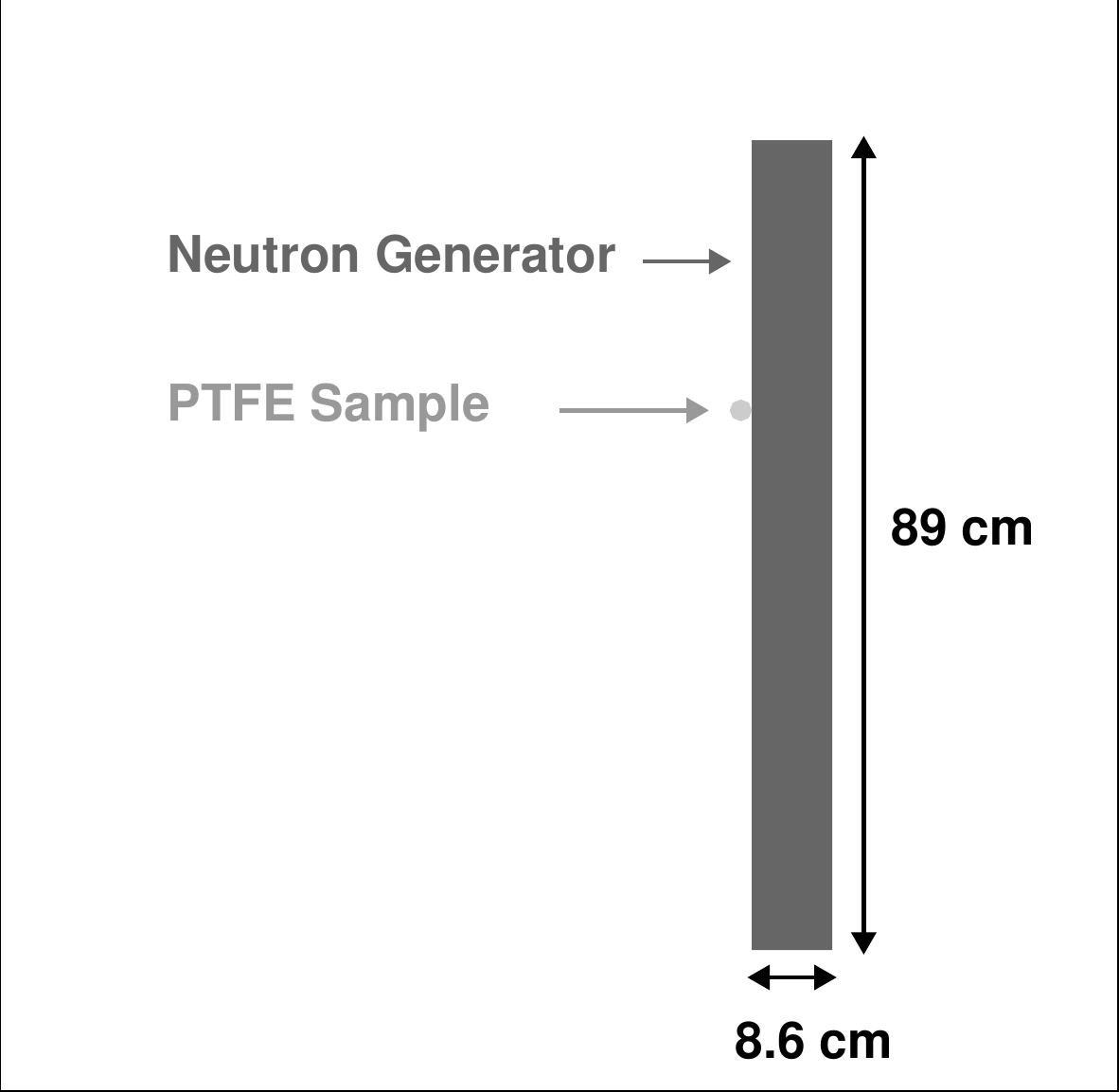}
\qquad
\includegraphics[width=.25\textwidth]{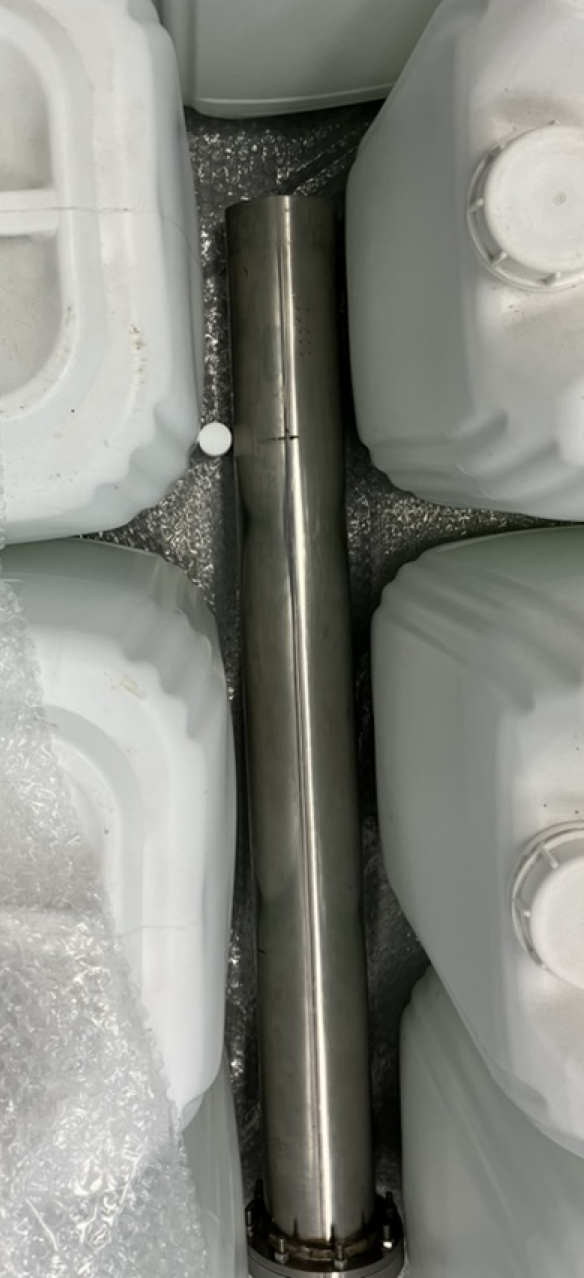}
\caption{
The left illustration is a schematic top view showing the arrangement of the deuterium-tritium neutron generator and the PTFE sample. 
The PTFE sample is positioned directly beside the neutron production point. 
The right photograph shows the actual setup. 
The white tanks surrounding the setup are water tanks used to slow down the neutrons from the neutron generator for safety.
}
\label{fig:F181}
\end{figure}
\begin{figure}[htbp]
\centering 
\includegraphics[width=.47\textwidth]{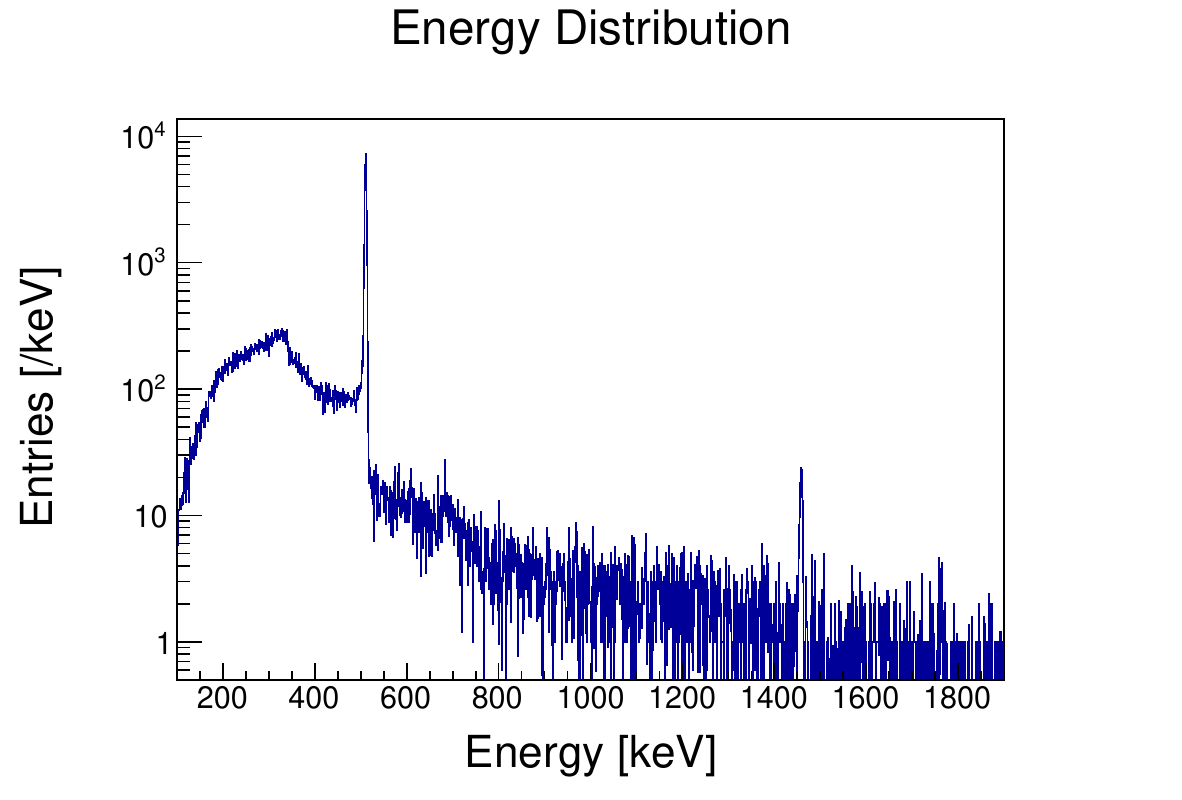}
\qquad
\includegraphics[width=.47\textwidth]{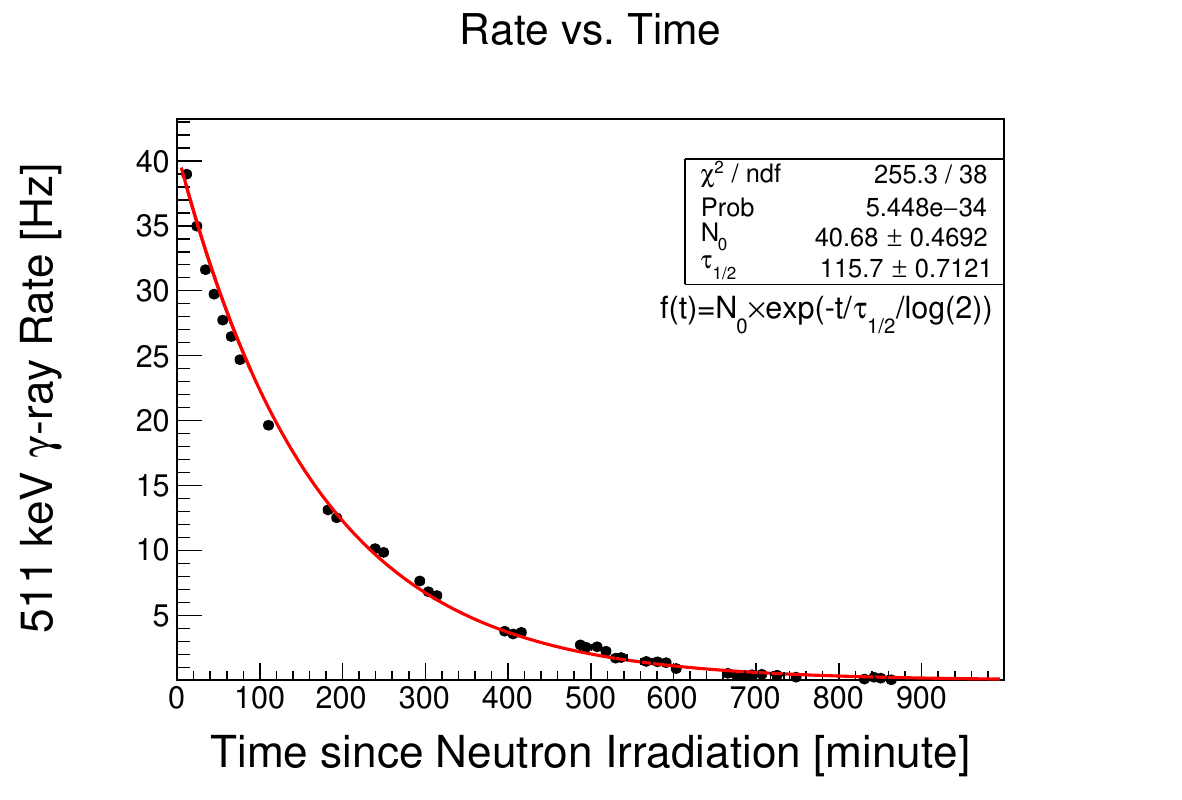}
\caption{
The left histogram shows the $\gamma$-ray energy spectrum obtained from the irradiated PTFE sample using the germanium detector at Shanghai Jiao Tong University. 
The peak at 511~keV corresponds to the absorption peak of the 511~keV $\gamma$-rays, which originate from positron annihilation produced by $\beta^{+}$-decay within the PTFE sample.
The lower energy components represent events where $\gamma$-rays deposited part of their energy in the detector. 
The peak near 1.46 MeV is attributed to $\gamma$-rays from $^{40}{\rm K}$, which is present in the surrounding environment. 
The right plot displays the time variation of the event rate around the 511~keV energy region. 
Since data acquisition with the germanium detector was occasionally interrupted, the time intervals between data points are not uniform.
By fitting the data with an exponential function, the half-life time of the event rate was determined to be approximately 116~minutes, which is roughly consistent with the expected $^{18}{\rm F}$ half-life time (110~minutes).
}
\label{fig:F182}
\end{figure}

As shown in Figure~\ref{fig:F181}, a PTFE sample (fluorine compound ${\rm C_{2}F_{4}}$) was placed near the neutron production point of the neutron generator.
High-energy neutrons were irradiated onto the sample, converting $^{19}{\rm F}$ within the PTFE into $^{18}{\rm F}$. 
We envision using PTFE as the $^{18}{\rm F}$ source since there is no risk of contaminating the detector even if it drops into the detector.
The PTFE sample is cylindrical, with a diameter of 2.4~cm and a height of 4.0~cm.
The ion source current in the neutron generator was set to approximately 65~$\mu$A, and the ion acceleration voltage was set to 85~kV.
After approximately 60 minutes of neutron irradiation, the production of $^{18}{\rm F}$ was verified by measuring the 511~keV $\gamma$-rays emitted from the PTFE sample using a high-purity germanium detector at Shanghai Jiao Tong University (GEM-25-76-HJ from ORTEC). 
The high-purity germanium detector did not have 4$\pi$ acceptance, and thus only one of the two 511~keV $\gamma$-rays was detected. 
The left panel of Figure~\ref{fig:F182} shows the $\gamma$-ray spectrum. 
A prominent peak corresponding to the 511~keV $\gamma$-ray is observed, with no other significant peaks detected in the nearby energy region. 
The radioactivity of the generated $^{18}{\rm F}$ was calculated based on the estimated $\gamma$-ray detection efficiency, which was determined by simulations considering the geometric conditions. 
The right panel of Figure~\ref{fig:F182} shows the time-dependent change in the detection rate of the 511~keV $\gamma$-ray after statistically subtracting background events from the surrounding environment, roughly consistent with the 110~minutes half-life of $^{18}{\rm F}$.

\subsubsection{Design of PTFE calibration source}
\label{subsubsec:f182}
\begin{figure}[htbp]
\centering 
\includegraphics[width=.55\textwidth, trim=2 2 2 2,clip]{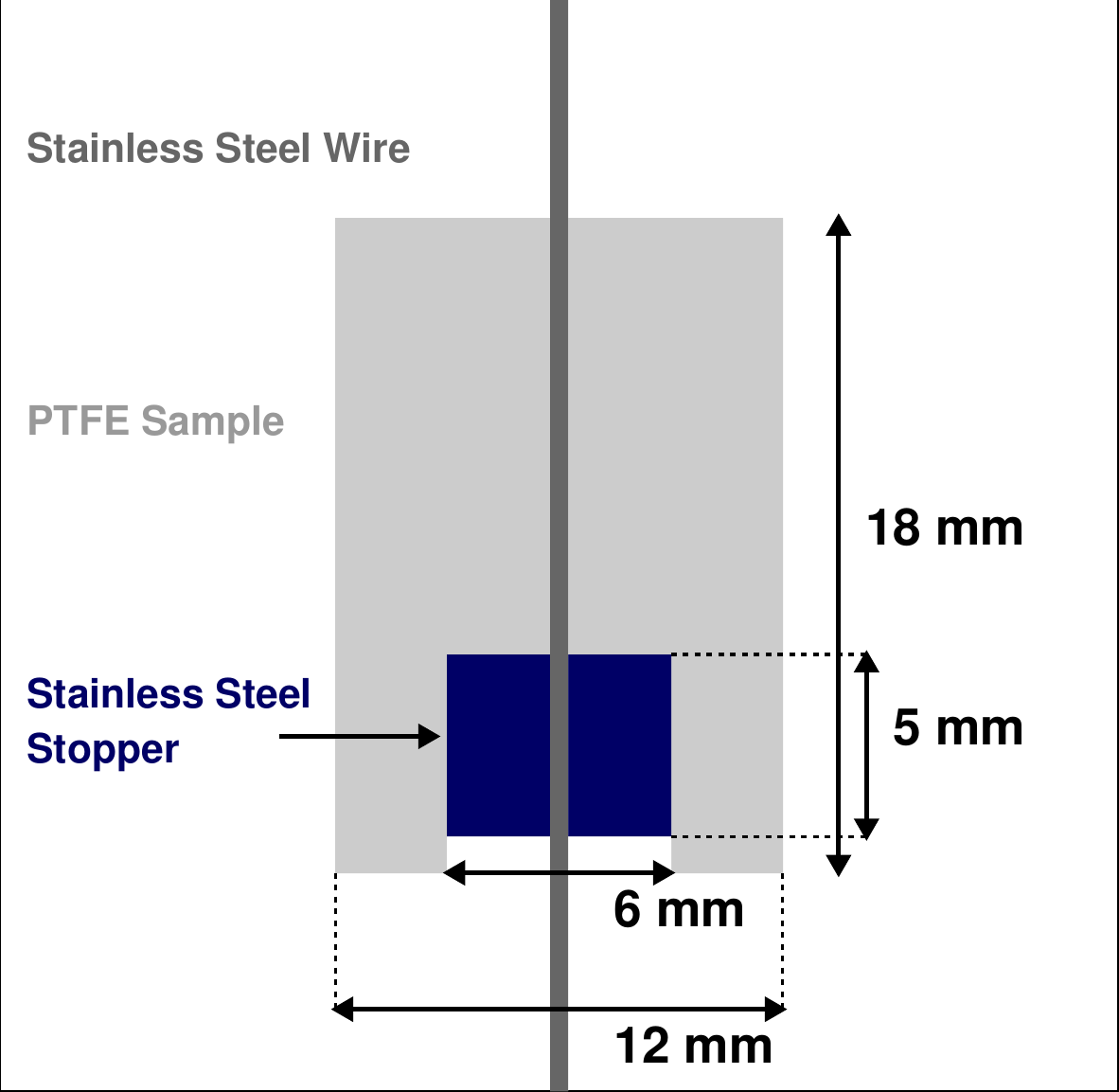}
\qquad
\includegraphics[width=.35\textwidth]{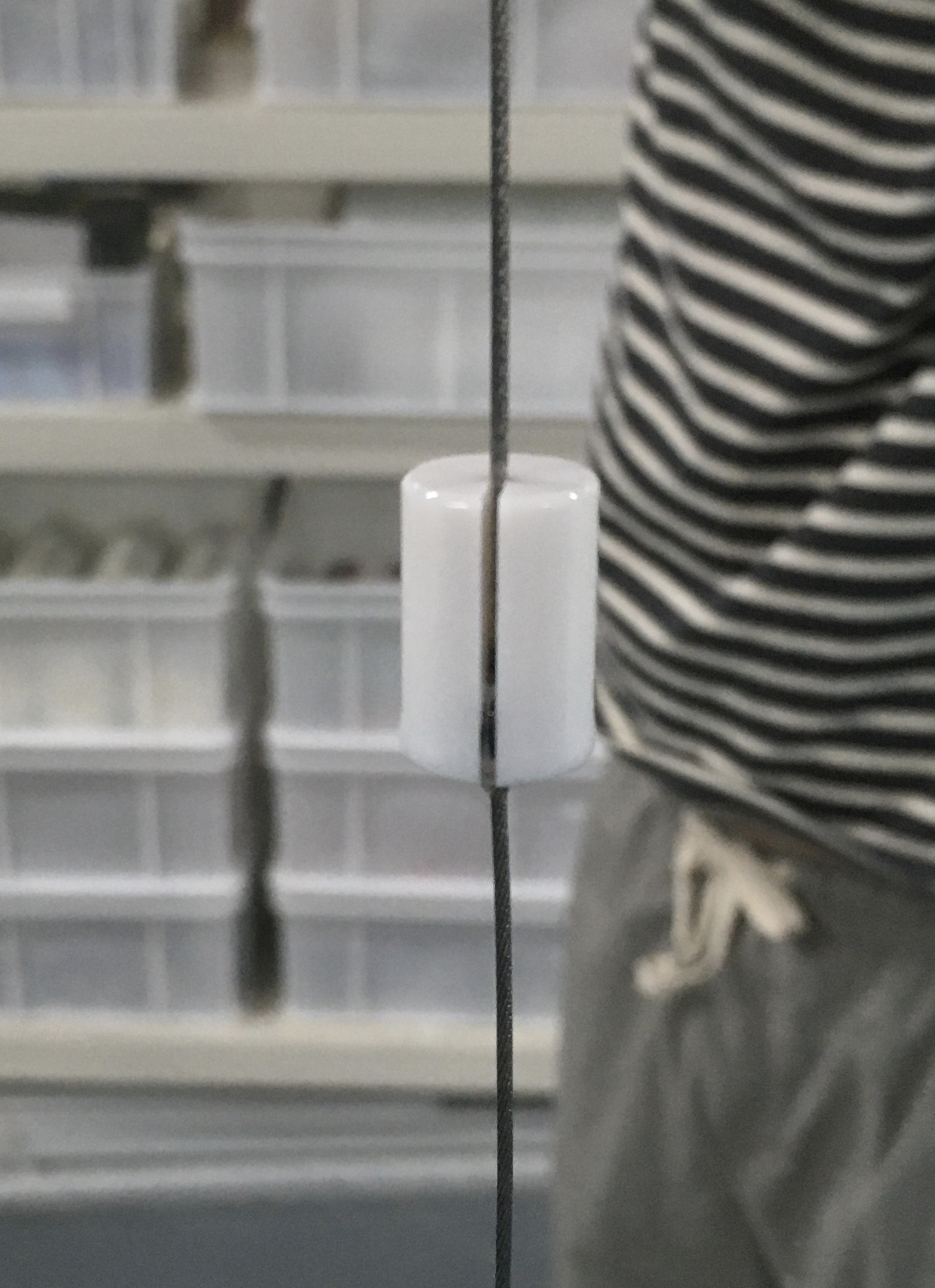}
\caption{
The left figure is a cross-sectional schematic diagram of the PTFE sample, the wire for attaching it, and the stainless steel stopper that fixes the PTFE sample position, which will be placed in the JUNO detector. 
The PTFE sample is cylindrical with a partially hollowed-out interior. 
The right photograph shows the PTFE sample, which has a slit that the stainless steel wire goes through.
The stainless wire is vertically fixed to the PTFE source by the stopper.
}
\label{fig:F183}
\end{figure}

The source radioactivity in JUNO should be around 100~Bq to optimize the data acquisition efficiency.
After the $^{18}{\rm F}$ generation test described above, it was found that increasing the ion source current of the deuterium-tritium neutron generator to 300~$\mu$A and the ion acceleration voltage to 100~kV increased the neutron generation rate by approximately six times compared to the previous test. 
As shown in Figure~\ref{fig:F183}, the PTFE sample to be placed in the JUNO detector will have an outer diameter of 1.2~cm and a height of 1.8~cm. 
It is estimated that an irradiation time of less than 10~minutes will be sufficient to activate a PTFE sample of this size into a 100~Bq $^{18}{\rm F}$ source. \par
The PTFE cylinder was designed with a slit and hollowed-out interior, allowing it to be directly attached to a stainless steel wire from the calibration source installation device using a stopper designed to prevent it from falling as shown in Figure~\ref{fig:F183}.
According to the JUNO detector simulation, the mean reduction in detected light in the two 511~keV $\gamma$-ray events produced by the $^{18}{\rm F}$ source, considering the shadow effect of the source itself and energy loss within the source, will be less than 1\%, meeting the experimental requirements.
We plan to generate the $^{18}{\rm F}$ radioactive source by irradiating this designed PTFE sample with fast neutrons using the deuterium-tritium neutron generator at the JUNO experimental site. 
The source will then be placed inside the detector using the calibration deployment system.

\subsection{Radioactive potassium-40 source}
\label{subsec:k40}
\begin{figure}[htbp]
\centering 
\includegraphics[width=.42\textwidth]{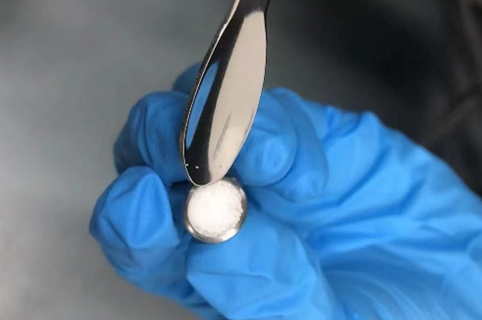}
\qquad
\includegraphics[width=.5\textwidth]{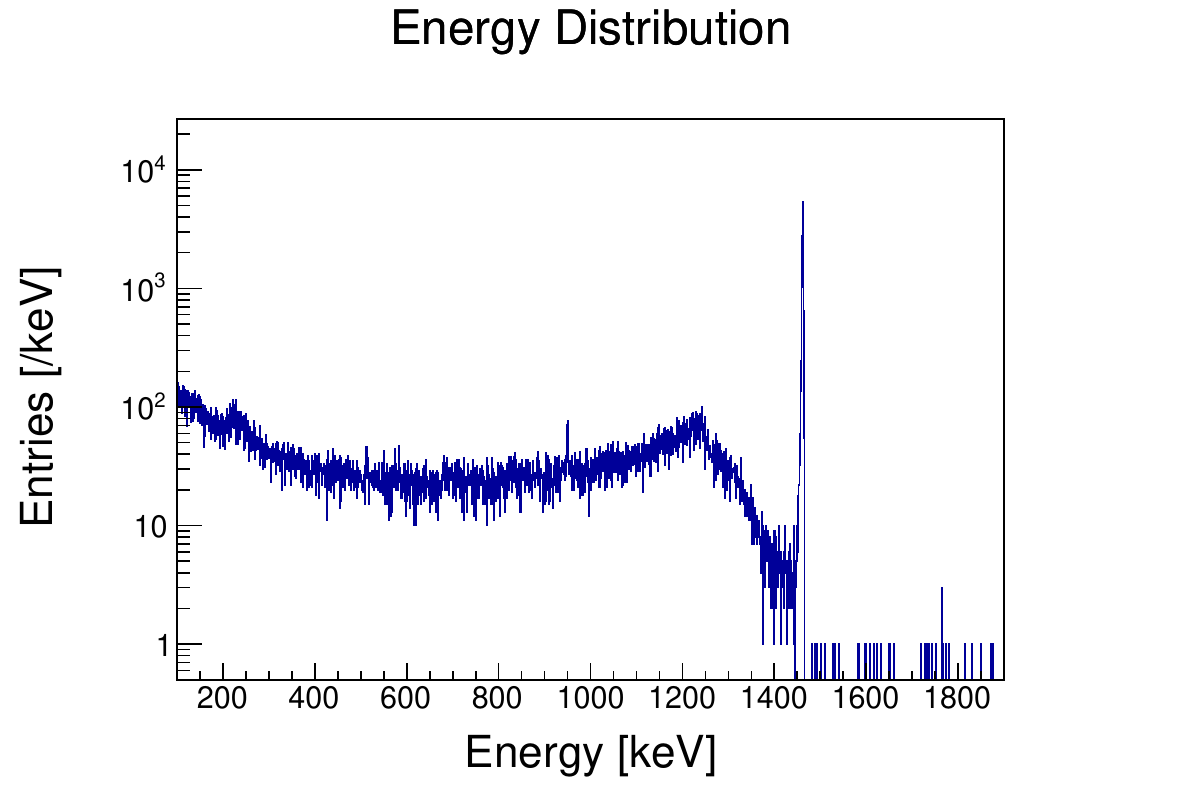}
\caption{
The left image shows the white potassium fluoride powder encapsulated in a titanium container. 
The right histogram displays the $\gamma$-ray energy spectrum measured from this source using a high-purity germanium detector at the China Jinping Underground Laboratory. 
The 1.46~MeV $\gamma$-ray peak was observed. 
The small peak around 950~keV corresponds to the so-called single escape peak.
}
\label{fig:K401}
\end{figure}

$^{40}{\rm K}$ is a long-lived radioactive isotope with a half-life of $10^{9}$~years~\cite{Chen:2017ngq}. 
Approximately 89\% of its decay leads to the production of calcium-40 $(^{40}{\rm Ca})$ via $\beta$-decay, while the remaining 11\% undergoes electron capture, resulting in the formation of argon-40 $(^{40}{\rm Ar})$ and the emission of a 1.46~MeV $\gamma$-ray. 
In this study, a radioactive calibration source was created by encapsulating potassium fluoride (KF) in a titanium (Ti) container. 
The left image of Figure~\ref{fig:K401} shows the potassium fluoride powder enclosed in the container. 
Since potassium-40 has a natural abundance of 0.01\%, and only 11\% of its decays emit $\gamma$-rays, the container was designed with a diameter of 12~mm and a height of 13~mm, approximately eight times larger by volume than other radioactive sources in JUNO~\cite{Zhang:2021tob}, to ensure sufficient radioactivity.
It was also confirmed that there was no corrosion of the container due to contact between the potassium fluoride and the titanium. \par

The $\gamma$-rays emitted from the sealed container were measured using a high-purity germanium detector at the China Jinping Underground Laboratory~\cite{PandaX-4T:2021lbm}. 
As shown in the right panel of Figure~\ref{fig:K401}, the 1.46~MeV $\gamma$-ray was clearly detected. 
Based on the estimated $\gamma$-ray detection efficiency from simulations, the radioactivity of $^{40}{\rm K}$ in the container was approximately 20~Bq (with about 11\% of this corresponding to $\gamma$-ray emission).
Since this radioactivity is still lower than that of other sources in JUNO, we plan to extend the calibration time for this source to collect a sufficient number of events.

\subsection{Radioactive radium-226 and americium-241 sources}
\label{subsec:ra226am241}
The following two radioactive sources were developed for the calibration of the low-energy region of the JUNO detector:
\begin{enumerate}
\item $^{226}{\rm Ra}$: $^{226}{\rm Ra}$ is an alpha-decay nuclide with a half-life of approximately 1,600~years~\cite{Balraj:2023tbf}. 
Upon decay, there is about a 4\% probability that a 186~keV $\gamma$-ray will be emitted from the de-excitation of radon-222 $(^{222}{\rm Rn})$.
Additionally, $\gamma$-rays are emitted during the decay of lead-214 $(^{214}{\rm Pb})$, bismuth-214 $(^{214}{\rm Bi})$, and lead-210 $(^{210}{\rm Pb})$, which are part of its decay chain. 
Among these, the $\gamma$-rays of 352, 295, 242, and 53.2~keV from $^{214}{\rm Pb}$~\cite{Zhu:2021qss} and 46.5~keV from $^{210}{\rm Pb}$~\cite{ShamsuzzohaBasunia:2014yyr} are expected to be useful for calibrating the low-energy region of the JUNO detector response.
\item $^{241}{\rm Am}$: $^{241}{\rm Am}$ is an alpha-decay nuclide with a half-life of approximately 433~years~\cite{Basunia:2005zz}. 
Upon decay, there is about a 36\% probability that a 59.5~keV $\gamma$-ray will be emitted from the de-excitation of neptunium-237 $(^{237}{\rm Np})$.
\end{enumerate}
The energies of some of the $\gamma$-rays mentioned here overlap with the background events caused by the $\beta$-decay of carbon-14 $(^{14}{\rm C})$ ($Q$-value of 156~keV) present in the liquid scintillator~\cite{ENSDF}. 
The contamination from $^{14}{\rm C}$ in the JUNO liquid scintillator is anticipated to be on the order of $10^{-17}$~g/g, which corresponds to approximately 40~kHz across the entire liquid scintillator in the JUNO detector. 
The effect from this background is expected to be mitigated by reconstructing the calibration event positions near the calibration source and by statistically subtracting background-only data taken when the source is removed from the detector~\cite{Takenaka:2024bya}. 
This paper does not delve into these analyses in detail but focuses on the preparation of the two radioactive sources for the JUNO experiment and the measurement of radioactivity using germanium ($\gamma$-ray detector) and silicon detectors (alpha-particle detector).


\subsubsection{Preparation of the radium-226 source}
\begin{figure}[htbp]
\centering 
\includegraphics[width=.35\textwidth]{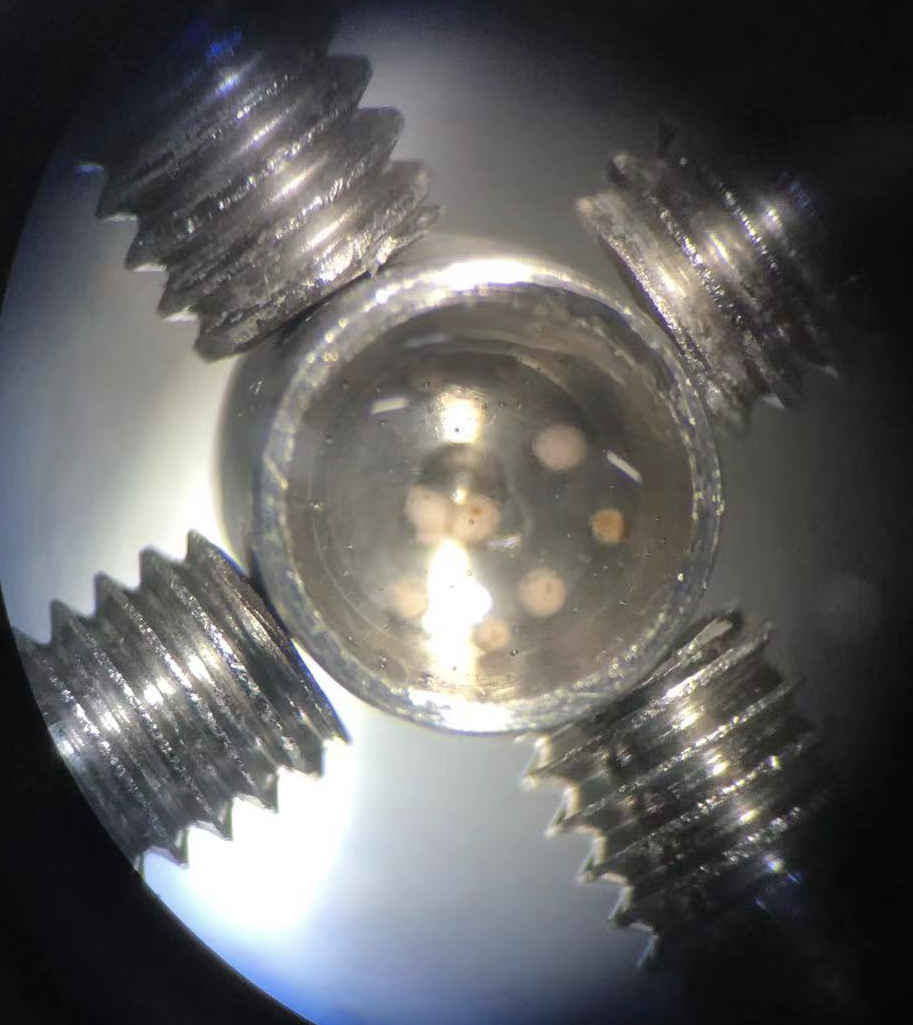}
\qquad
\includegraphics[width=.55\textwidth]{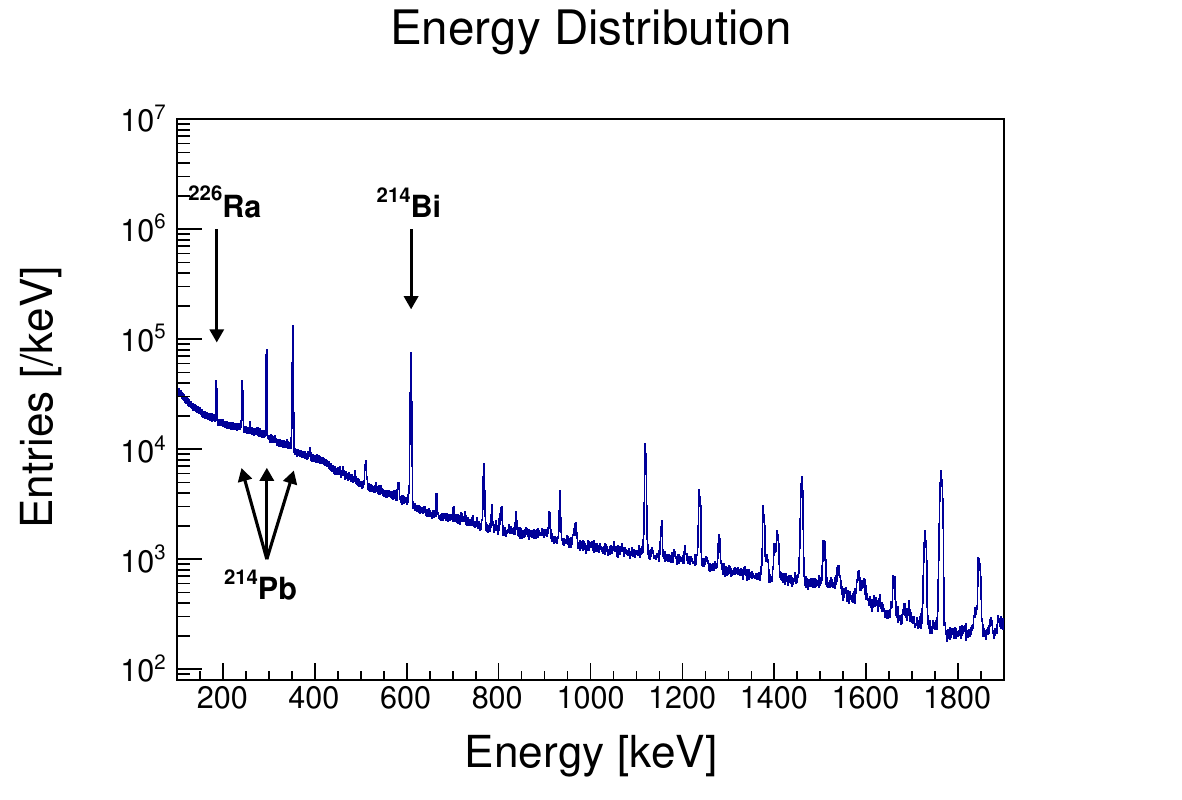}
\caption{
The left photograph shows the interior of the radium source container, captured using a microscope. 
The granular objects are small particles that adsorbed $^{226}{\rm Ra}$, sealed together with epoxy gel. 
The right histogram displays the $\gamma$-ray energy spectrum measured from this source using a germanium detector. 
$\gamma$-rays of 186~keV from the decay of $^{226}{\rm Ra}$, 352, 295, and 242~keV from the decay of $^{214}{\rm Pb}$, and 609~keV from the decay of $^{214}{\rm Bi}$ are clearly observed. 
$^{214}{\rm Bi}$ decay produces $\gamma$-rays of various energies besides the 609~keV line, such as 1847, 1764, 1730, 1408, 1378, 1238, 1120, 806~keV and so on.
Other smaller peaks correspond to these energies.
}
\label{fig:Ra2261}
\end{figure}

At Nanhua University, $^{226}{\rm Ra}$ was adsorbed onto particulate porous materials, and these small particles were subsequently provided.
These small particles were encapsulated in a titanium container with a diameter and height of 6~mm using the encapsulation method, which will be introduced in Section~\ref{subsec:sourceweld}. 
The left image in Figure~\ref{fig:Ra2261} shows a photograph of the interior of the source container, where the $^{226}{\rm Ra}$-adsorbed particles are sealed together with epoxy gel.
After encapsulation, $\gamma$-rays from the source were measured using the germanium detector at Shanghai Jiao Tong University. \par

The right plot in Figure~\ref{fig:Ra2261} displays the $\gamma$-ray energy spectrum measured by the germanium detector, where the 186~keV $\gamma$-rays from $^{226}{\rm Ra}$ and three $\gamma$-rays from $^{214}{\rm Pb}$ were observed. 
The radioactivity per one small particle, corrected for the $\gamma$-ray detection efficiency obtained from simulations based on Geant4, was determined to be approximately 40~Bq with $\pm10$~Bq variations.
By varying the number of small particles encapsulated, three sources with different levels of radioactivity were prepared. 
The estimated radioactivities obtained from the germanium detector for each of these sources were 140, 570, and 970~Bq. 
Based on the actual $^{14}{\rm C}$ event rate in JUNO and the data acquisition efficiency, the most suitable source among these three will be selected for detector calibration.

\subsubsection{Preparation of the americium-241 source}
\begin{figure}[htbp]
\centering 
\includegraphics[width=.35\textwidth]{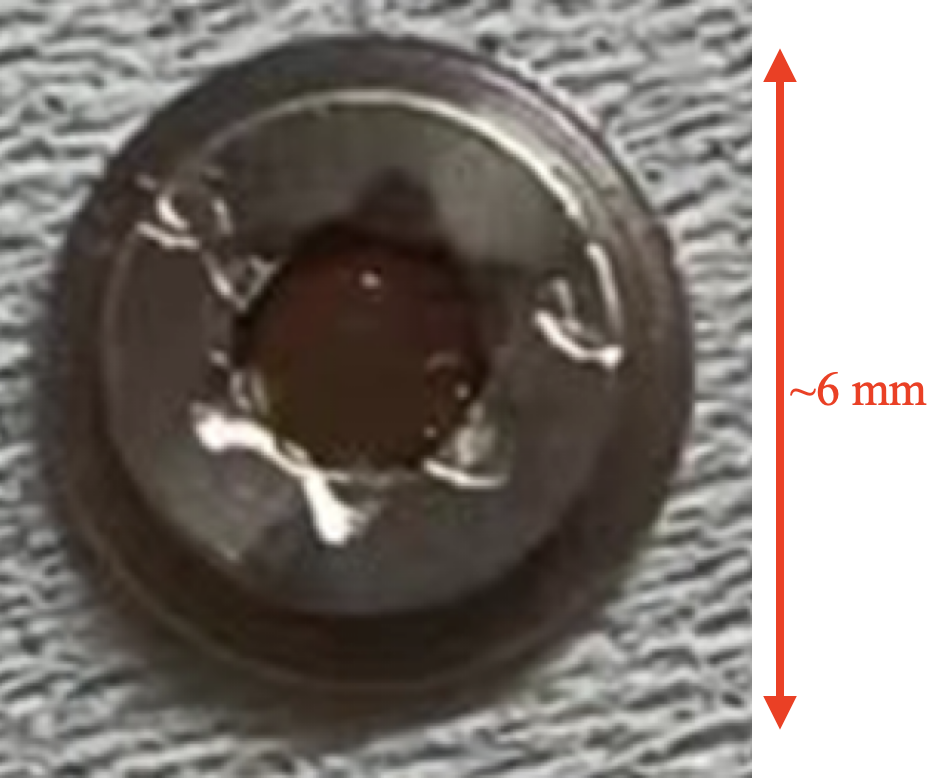}
\qquad
\includegraphics[width=.55\textwidth]{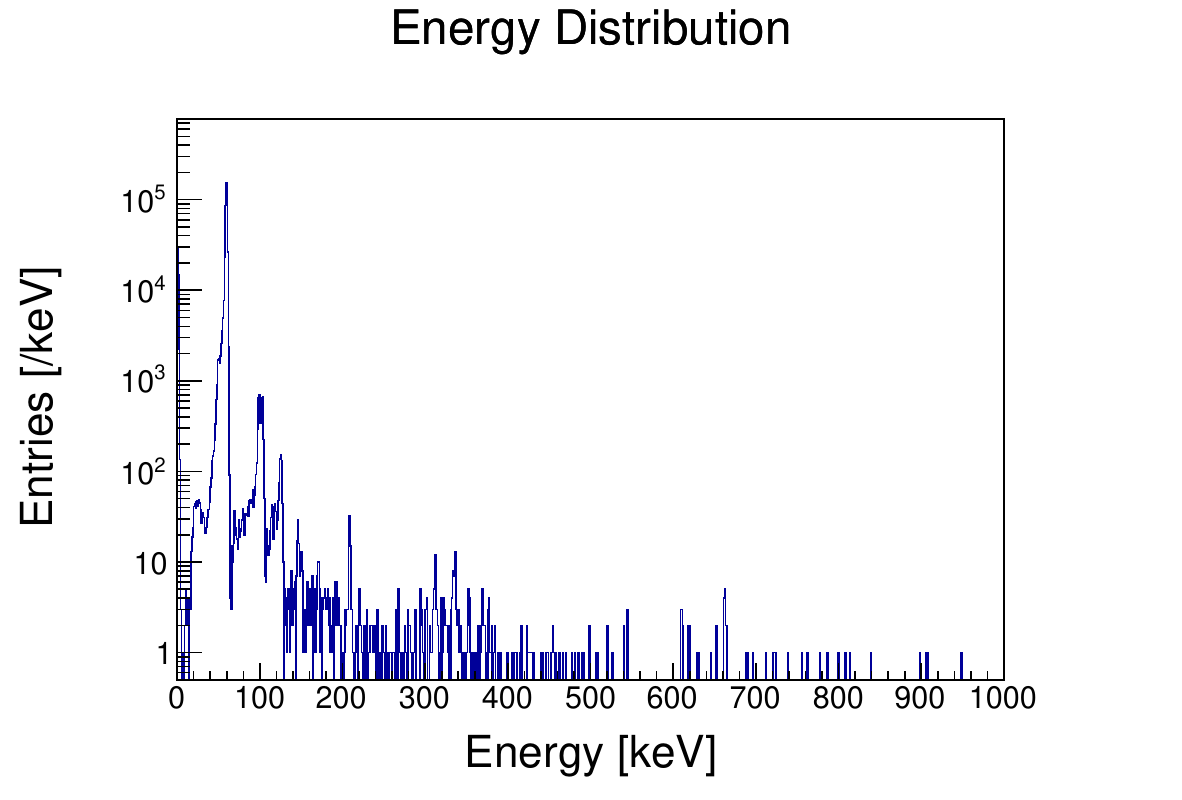}
\caption{
The left image shows a photograph of the americium source extracted from a smoke detector. 
It is covered with protective metal and positioned at the center, with an outer diameter, including the protective cover, of approximately 6~mm. 
The right histogram displays the $\gamma$-ray energy spectrum obtained using the high-purity germanium detector at the China Jinping Underground Laboratory. 
A 59.5~keV peak associated with alpha decay and other $\gamma$-ray peaks, such as those around 100 and 125~keV, with lower emission branching ratios from $^{241}{\rm Am}$ were observed. 
}
\label{fig:Am2411}
\end{figure}

An $^{241}{\rm Am}$ source, commonly used in commercial smoke detectors, was extracted and encased in a source container to serve as a calibration source for the JUNO experiment. 
The left image in Figure~\ref{fig:Am2411} shows the $^{241}{\rm Am}$ source extracted from the smoke detector. 
To measure the radioactivity of the source, the emission rate of the alpha-particle from the source was measured using a silicon detector (Alpha Mega from ORTEC). 
After applying a detection efficiency correction based on simulations, the radioactivity was estimated to be approximately 6~kBq per $^{241}{\rm Am}$ source. 
Additionally, the 59.5~keV $\gamma$-rays emitted after alpha decays were measured using a high-purity germanium detector at the China Jinping Underground Laboratory. 
These germanium detectors allowed for $\gamma$-ray measurements with reduced background interference from the surrounding environment. 
As a result, a distinct $\gamma$-ray peak at 59.5~keV was observed, as shown in the right plot of Figure~\ref{fig:Am2411}.
Similar to the $^{226}{\rm Ra}$ source, this $^{241}{\rm Am}$ source was also encased in a titanium container with a diameter of 10~mm and a height of 6~mm using the encapsulation method discussed in detail later.

\subsection{Source sealing method}
\label{subsec:sourceweld}
This section introduces the welding techniques employed to encapsulate and seal the $^{40}{\rm K}$, $^{226}{\rm Ra}$, and $^{241}{\rm Am}$ sources within titanium containers. 
A key challenge in sealing radioactive source containers through welding is the risk of source evaporation due to the heat generated during the welding process. 
In this study, measures were taken to minimize the generated heat or to prevent it from being transferred to the radioactive source. 
These measures will be discussed in Section~\ref{subsubsec:sourceweld1}.
Additionally, Section~\ref{subsubsec:sourceweld2} will address the assessment of the seal quality and the impact of the welding process on the radioactive sources.

\subsubsection{Low-temperature welding technique}
\label{subsubsec:sourceweld1}
\begin{figure}[htbp]
\centering
\includegraphics[width=0.75\linewidth]{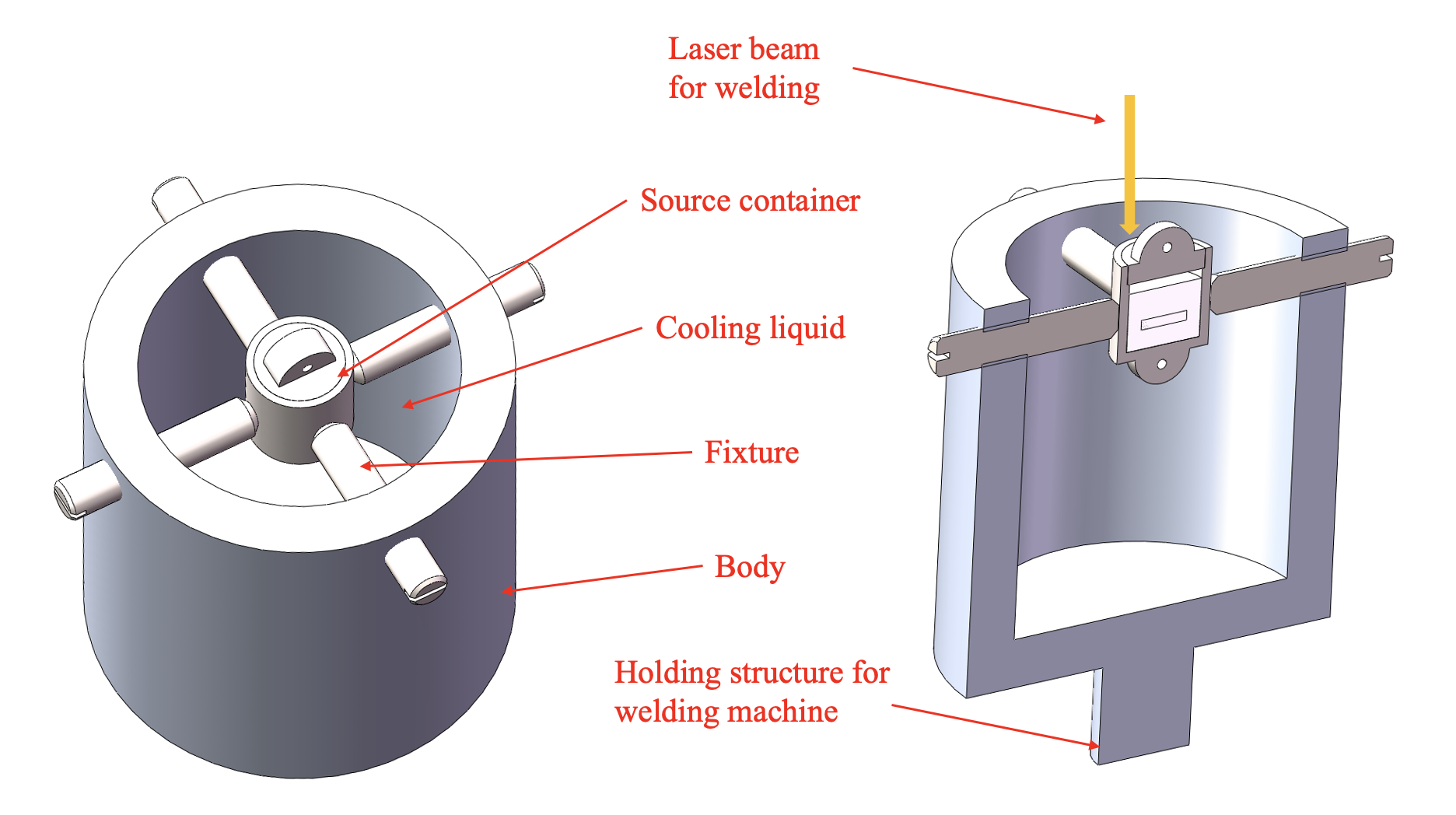}
\caption{
A schematic diagram of the fixture used during laser welding to secure the source container position. 
The fixture consists of structures (``Fixture'') that hold the source container in place and a larger vessel (``Body'') for storing the cooling liquid.
}
\label{fig:sourceweld1}
\end{figure}

\begin{figure}[htbp]
\centering
\includegraphics[width=0.75\linewidth]{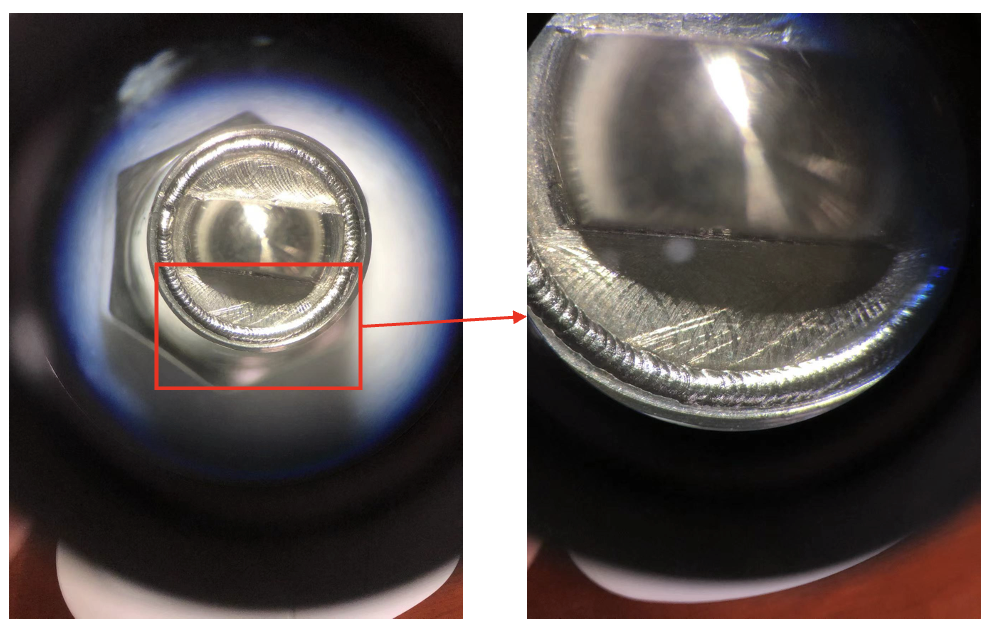}
\caption{
Photographs of the titanium container after welding. 
The image on the right is a close-up of the welded section.
}
\label{fig:sourceweld2}
\end{figure}

To minimize the impact of heat generated during welding on the radioactive sources, the following three measures were implemented:
\begin{enumerate}
\item Welding technique: Laser welding was adopted in this study instead of arc welding. 
This is because laser welding generates less heat than arc welding, making it more suitable for this application.
\item Inclusion of epoxy gel: For the welding of $^{226}{\rm Ra}$ and $^{241}{\rm Am}$ source containers, as shown in the left image of Figure~\ref{fig:Ra2261}, the epoxy gel was filled inside the container to seal the sources and to reduce heat transfer.
\item Cooling liquid: A fixture was designed, as illustrated in Figure~\ref{fig:sourceweld1}, to secure the source container while keeping it in contact with a liquid stored at the bottom. 
This arrangement allowed the heat generated during welding to be effectively removed, with the liquid functioning as a coolant for the source container.
In this study, room-temperature water was used as the coolant during the welding process.
\end{enumerate}
With these measures, the source containers were sealed using laser welding. 
Figure~\ref{fig:sourceweld2} shows a photograph of the titanium container after welding.

\subsubsection{Quality check}
\label{subsubsec:sourceweld2}
After laser welding, the sealing quality of the source container and the absence of any loss of the radioactive source were verified through the following three tests:
\begin{enumerate}
\item To check for any leaks, the welded container was placed in a chamber capable of filling with helium gas up to 1~MPa for half a day. 
If there were any leaks, helium gas would leak into the container. 
The source container was then transferred to another vacuum chamber connected to a helium detector (PHOENIX Quadro from Leybold) to measure the possible helium signal from the source container. 
The measurement results showed no significant helium detection (with an upper leakage rate limit of $10^{-11}$~mbar$\cdot$L/s).
\item If there were leaks, there could be radioactive contamination on the outer surface of the container. 
To check for it, the welded container was immersed in pure water for two days, and then the $\gamma$-rays from the water were measured using a high-purity germanium detector. 
No significant difference was observed compared to a blank water sample within the statistical uncertainty of the measurement, confirming that the radioactive source on the external surface of the container is less than 0.1\% of the radioactive source inside.
\item The radioactivity of the source inside the container was measured before and after welding using a high-purity germanium detector. 
No significant change in radioactivity was observed within an uncertainty range of approximately 10\%.
The 10\% uncertainty originates from the reproducibility of the radioactive source and its container position relative to the germanium detector, estimated by simulation.
\end{enumerate}
These tests confirm that the source container produced in this study can be safely introduced into the JUNO detector.

\section{Conclusion}
\label{sec:conclusion}

We customized the laser system and radioactive sources for use in the JUNO experiment. 
The design of the diffuser ball, which diffuses laser light within the detector, was optimized to ensure that the photon emission time across the surface of the ball remains uniform within a range of $\pm0.25$~nsec.
Additionally, we introduced a tuner that uses optical filters to adjust the light intensity over a range of four orders of magnitude.
We conducted tests on the $^{18}{\rm F}$ source, generated by irradiating PTFE with fast neutrons, confirming the production of $^{18}{\rm F}$ through the observation of 511~keV $\gamma$-rays and the measurement of the decay constant. 
Based on these results, we determined the shape of the PTFE sample to be used for JUNO detector calibration. 
Moreover, we prepared $^{40}{\rm K}$, $^{226}{\rm Ra}$, and $^{241}{\rm Am}$ sources and verified their sufficient radioactivity using germanium and silicon detectors. 
The $^{40}{\rm K}$ source was created from potassium fluoride powder, and by enlarging the source container, we achieved a radioactive source of 20~Bq. 
Considering the impact of background events from $^{14}{\rm C}$ in the low-energy range in the JUNO detector, we prepared radium and americium sources with activities ranging from 100~Bq to several~kBq. 
Using a low-temperature welding technique, we sealed these radioactive sources into their containers.
All of the customized calibration sources in this study will be placed into the JUNO detector with the calibration source deployment devices.

\acknowledgments
The authors gratefully acknowledge Shiwei Jing at Northeast Normal University for assistance with the use of the neutron generator, Shoukang Qiu at Nanhua University for providing the $^{226}{\rm Ra}$ source, and Dalin Fu for the support with the welding work. 
We also appreciate all of the JUNO collaborators who gave us helpful comments and members at Shanghai Jiao Tong University and the China Jinping Underground Laboratory for supporting our work. \par
This work is supported by the National Key Research and Development Program of China (Grant number: 2023YFA1606104), National Science Foundation of China (Grant number: 12250410235, 12222505), and the Strategic Priority Research Program of the Chinese Academy of Sciences (Grant number: XDA10010800).
Y. M. and J. H. thank the sponsorship from the Yangyang Development Fund.

\bibliographystyle{JHEP} 
\bibliography{reference}

\end{document}